\documentclass[aps,prl,showpacs,superscriptaddress]{revtex4-1}
\usepackage{amsmath}
\usepackage{epsfig}
\usepackage{amssymb}
\usepackage{dcolumn}
\usepackage{graphicx}
\usepackage{dcolumn}

\usepackage{color}

\newcommand{\beq}{\begin{equation}}
\newcommand{\eeq}{\end{equation}}
\newcommand{\dpsi}{\delta \psi}
\newcommand{\rv}{\mathbf{r}}
\newcommand{\vv}{\mathbf{v}}
\newcommand{\kv}{\mathbf{k}}
\newcommand{\pv}{\mathbf{p}}
\newcommand{\qv}{\mathbf{q}}

\begin{document}
\date{\today}

\title{Effects of the non-parabolic kinetic energy on non-equilibrium polariton condensates}

\author{F. Pinsker}
\affiliation{Clarendon Laboratory, Department of Physics, University of Oxford, Parks Road, Oxford OX1 3PU, United Kingdom.}
\email{florian.pinsker@physics.ox.ac.uk }

\author{X. Ruan}
\affiliation{Department of Mathematics, National University of Singapore, Singapore.}

\author{T. J. Alexander}
\affiliation{School of Physical, Environmental and Mathematical Sciences, UNSW Canberra, Canberra ACT 2600, Australia.}


\begin{abstract}  In the study of non-equilibrium polariton condensates it is usually assumed that the dispersion relation of polaritons is parabolic in nature.  We show that considering the true non-parabolic kinetic energy of polaritons leads to significant changes in the behaviour of the condensate due to the curvature of the dispersion relation and the possibility of transfer of energy to high wavenumber components in the condensate spatial profile.  We present explicit solutions for plane waves and linear excitations, and identify the differences in the theoretical predictions between the parabolic and non-parabolic mean-field models, showing the possibility of symmetry breaking in the latter. We then consider the evolution of wavepackets and show that self-localisation effects may be observed due to the curvature of the dispersion relation.  Finally, we revisit the dynamics of dark soliton trains
and show that additional localized density excitations may emerge in the dynamics due to the excitation of high frequency components, mimicking the appearance of near-bright solitary waves over short timescales.


\end{abstract}

\maketitle

\section{Introduction}   Light-matter coupling for microcavity photons and excitons leads to two branches of quasiparticles, known as exciton-polaritons, which at low enough densities may undergo Bose-Einstein condensation (see e.g. \cite{KeelingCP2011,Light} for discussions of the condensation process).  Most mean-field approaches for the polariton condensate have used a parabolic approximation for the lower polariton dispersion profile, so yielding a complex Gross-Pitaevskii equation (cGPE) as a model for the lowest energy polariton mode \cite{keen, keeli, Wout}.  The lower polariton mode can be distinctively addressed due to the large energy gap between the two branches \cite{ke}, and a constant `effective mass' is then typically invoked to explore the condensate properties \cite{Light,keeli, Wout}. 
Recently the constant effective mass approximation in polariton condensates has been relaxed to include {\it velocity dependent} effective mass effects in novel dynamical wave models \cite{mk, mk2}.  This has allowed the development of a mean-field theory for the coherently driven polariton condensate wave function \cite{mk}.  Here we take a step further and incorporate a realistic dispersion relation for the polaritons, which leads to a marked deviation from the usual Gross-Pitaevskii theory \cite{leg, revBEC}. We note that this approach has also been taken recently in another work \cite{WinklerPRB2016}, however here we explicitly compare the different approaches and identify some of the new features which may emerge from incorporating the full dispersion profile in the theoretical model.  It will be shown that the effective mass concept still naturally emerges within this generalisation of the state equation.  While numerous experiments \cite{alex, PRL, PRX} show good agreement between the classic (parabolic) cGPE and experimental measurements, for observations at larger wave vectors $k$, predictive deviations between the models discussed here are to be expected and so care has to be taken when seeking a theoretical explanation.

Our paper is structured as follows: first we introduce the mean-field models under consideration; we then state analytical results for these mean-field models; next we examine the localisation mechanism of polaritons based on the non-parabolic dispersion relation, and finally we revisit some dark soliton phenomena and indicate differences between the models.

\subsection{Polariton condensate models and mean-field theory assumptions}

Before we turn to the different semi-classical polariton condensate models we note that the underlying polariton many-body theory contains several approximations  \cite{keeli}. The interaction of photons with electrons and holes makes use of the dipole and the rotating wave approximation. The Hamiltonian of the polariton modes \cite{flaa,east,keeli} has an effective $k$ dependent interaction strength due to the variation of the Hopfield coefficients \cite{Hopf} in $k$ along the lower/upper polariton branch (see e.g. \cite{Light}), the Coulomb interaction becomes stronger as the polariton becomes more excitonic, and the saturation interaction is strongest when the polariton consists equally of photon and exciton components.  We assume that the scattering length is much shorter than the de Broglie wavelength, and we do not preserve the effect of Hopfield coefficients on the interaction as it would take into account the decrease of both Coulomb and saturation effects for large exchanged momenta \cite{keeli}. However we do include the polarisation of polaritons, so the polariton condensate spinor wave function $\boldsymbol{\psi}=(\psi_+,\psi_-)$ for the lowest energy mode is governed by two coupled complex Ginzburg-Landau-type partial differential equations \cite{keen, Wout,  Light, Rev1}, which take into account the effects of weakly short-range and polariton-reservoir interactions, as well as non-equilibrium properties such as: incoherent gain and decay of condensate polaritons; energy relaxation \cite{relax}; and spin/polarisation \cite{spin, PRX,PRL}. Note that whenever we consider the spin coherent case $\psi_+ = \psi_-$ we use the notation $\psi$ for the condensate wave function. Due to a constant loss of information, corresponding to the rapid decay of polaritons, the condensate is effectively semiclassical (as observed experimentally e.g. in \cite{alex, PRX, PRL}). For the more general quantum treatment and implications however we refer to \cite{ keeli,east,flaa}.  In addition we take into account the fact that the {\it kinetic energy is non-parabolic} \cite{Light}, i.e. it varies due to the $k$ dependence of the Hopfield coefficents.  We define the dispersion relation (or equivalently, kinetic energy density), of the lower (upper) polariton branch as \cite{Light, keeli}, 
\begin{equation}
\label{dispersion}
 \omega_{_{\rm L,U}} (\kv) = \frac{1}{2} \big(\omega_{\rm cav} (\kv) +
\omega_{\rm exc} (\kv)   \mp \sqrt{(\omega_{\rm cav}
(\kv)-\omega_{\rm exc} (\kv))^2 + 4  \Omega^2_R} \big).
\end{equation} 
Note that the lower and upper polariton modes can be distinctively addressed experimentally due to the large energy gap between the two branches (see Fig. \ref{F1}).
Here the dispersion of the cavity photon is $\omega_{\rm cav} (k) = \frac{c}{n_0} \sqrt{q_z^2 + k^2}$.  We have used the notation $k=|\kv|$ for $\kv\in {\mathbb R}^D$ with $D=1,2$, with $c$ denoting the speed of light, $q_z=\frac{2\pi M}{l_z}$ the quantization of the confined photon in the $z$-direction, $M$ 
the number of the quantized $z$-mode orthogonal to the $\kv$-plane,
$n_0$ the refractive index between the cavity mirrors, and $l_z$  the cavity spacer length.
The dispersion of the exciton $\omega_{\rm exc} (\kv) \approx \omega_{\rm exc}^0 + 10^{-4} k^2 \approx \omega_{\rm exc}^0$ can be assumed to be constant close to the centre of the polariton dispersion.
 The minimum splitting
between the two dispersions, which is obtained at $\omega_{\rm cav}(\kv)=\omega_{\rm exc}(\kv)$, is given by $2\Omega_R$.
We set $\hbar\omega_{\rm exc}^0 = 1.557$ eV,
the mass of the cavity photon $m_{\rm cav} \sim  10^{-4}-10^{-5} m_e$
and the effective exciton mass $m_{\rm exc} \sim 0.1-1 m_e$ with $m_e$ the electron mass,
reflecting the experimental values used in Refs. \cite{PRX,PRL}. While we reduce our considerations to the mean-field states we note for the sake of completeness that the polariton can be described quantum mechanically by the Hamiltonian \cite{flaa}
\beq
\hat {\mathcal H} = \hat {\mathcal H}_{\rm kin} + \hat {\mathcal H}_{\rm p-p} + \hat {\mathcal H}_{\rm p-ph} + \hat {\mathcal H}_{\rm pump},
\eeq
where the kinetic energy of the polariton mode generated by $\hat a_{\kv}$ and the potential energy due to elastic polariton-polariton scattering are given by
\beq
\hat {\mathcal H}_{\rm kin}  = \sum_{\kv}  \omega_{_{\rm L,U}} (\kv) \hat a^\dagger_{\kv} \hat a_{\kv} \hspace{10mm}  \hat {\mathcal H}_{\rm p-p}  = \sum_{\kv_1, \kv_2, \pv}  U_{\kv_1 \kv_2 \pv} \hat a^\dagger_{\kv_1} \hat a^\dagger_{\kv_2} \hat a_{\kv_1 + \pv}  \hat a_{\kv_2 - \pv} 
\eeq
correspondingly and $U_{\kv_1, \kv_2, \pv} = U_0 X_{\kv_1} X_{\kv_2} X_{\kv_1+\pv}X_{\kv_2 - \pv}$, where $U_0 =\frac{6E_b a^2_B}{S}$, $E_b$ and $a_B$ are the exciton binding energy and
the Bohr radius, and $S$ is the system area. The incoherent processes for interactions of polaritons with a thermal bath  of acoustic phonons and the effect of incoherent pumping in the rotating wave approximation is given by \cite{flaa}
\beq
\hat {\mathcal H}_{\rm p - ph}  = \sum_{\kv_1, \kv_2} \int  \frac{L_z}{2 \pi} d q_z G_\qv \hat a^\dagger_{\kv_1} \hat a_{\kv_2} \hat b_{\qv} + h.c.\hspace{10mm}  \hat {\mathcal H}_{\rm pump}  = \hbar \sum_{\kv \eta}   \left( g_{\kv \eta} \hat a_{\kv} \hat d^\dagger_{\eta} + g^*_{\kv \eta} \hat a^\dagger_{\kv} \hat d_{\eta} \right),
\eeq
with $\hat b_{\qv}$ being the phonon destruction operators and $\hat d_{\eta}$ the operators corresponding to the bosonic pumping reservoir, while we refer to \cite{flaa} for further details about the functions $L_z$, $G_\qv $ and $g^*_{\kv \eta}$ and limitations of this approach. Now supposing the above assumptions and the quantum state to be in a product state and that a complex mode $\psi$ carries all relevant information of the many-body system we directly obtain a mean-field partial differential equation \cite{Light}. Here in particular the kinetic energy becomes
\beq
\hat {\mathcal H}_{\rm kin}  \to  \int \omega_{_{\rm L,U}} (|\kv|) |\psi (\kv)|^2,
\eeq
whose nature and implications are the topic of this paper. The interaction terms are very small under state-of-the-art experimental conditions and the interactions between the condensed phase and the environment  and pumping are modelled in polariton mean-field theory phenomenologically \cite{Light, Rev1}.

While the kinetic energy density is the ``complete" factor for calculating the kinetic energy of the polariton condensate, the effective mass has played a significant role in recent publications. Next we clarify its nature.

\section{Results}

\subsection{Approximations to the kinetic energy}

By looking at the kinetic energy density, the effective mass $m(k)$ of particles in semiconductors can be seen to depend on the curvature of the particles' dispersion relation $\omega(k)$ at a particular $k$-value \cite{Dresselhaus,mk,Ober}:
\beq
\frac{k^2}{2 m  (\kv)} := \frac{k^2 \omega^{\prime\prime}_{\rm L,U}(k)}{2},
\eeq
where $f(k) ''$ means second derivative of $f$ in $k$ (i.e. curvature with respect to $k$).  At the simplest level of approximation, the curvature of the dispersion relation is taken to be fixed, as is the case for a free particle with its associated parabolic dispersion.  For the case of a polariton condensate this approximation is valid for small values of $k$.  This approximation is widely used in polariton condensate modelling, leading to a second derivative/Laplacian for the dispersion in the equations of motion.  The next level of approximation allows for the effective mass to vary with $k$, i.e. a ``velocity dependent effective mass", as has been used for instance in Ref. \cite{mk}.  This approach effectively reduces the dispersion relation to the quadratic term in the Taylor series expansion of the full dispersion, with some constant offset.  Finally, the third approach, and the one we focus on, is to use the full dispersion in the equations of motion.

\begin{figure}[ht]
\vspace{20mm}
\begin{tabular}{c}
\begin{picture}(150,0)
\put(-210,-80) {\includegraphics[scale=0.2]{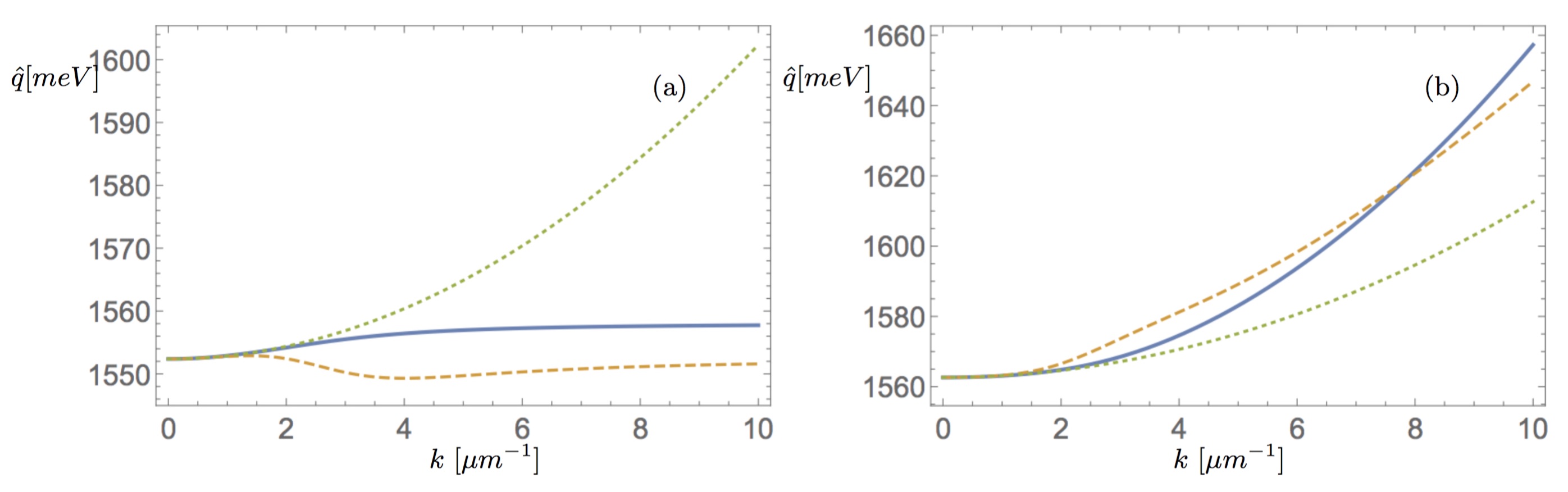} }
\end{picture} 
\end{tabular}

\vspace{25mm}

\caption{ (a) Energy dispersions of the lower polariton branch: Solid line corresponds to the lower polariton dispersion, dashed line to the velocity dependent effective mass model \cite{mk} and dotted line to parabolic approximation with energy shift. (b) Dispersions of the upper branch: The lines are associated as before.}
\label{F1}
\end{figure}

Note that while we consider here a single site, in contrast to lattices of polariton condensates, the semiconductor itself possesses a lattice structure which acts on the electrons and holes, i.e. the effective mass is a product of the polariton mean-field model. 
Due to this effective mass the kinetic energy of a polariton condensate varies with $k$ non-parabolically. This leads to the approximate approach to the mean-field description considering the kinetic energy to be  
 \beq
 E^{\rm m(k)}_{\rm kin} = \int \bigg( \frac{|\kv|^2}{2 m  (\kv)} + \omega_{\rm i} \bigg) |\psi (\kv)|^2 \equiv  \int \hat q^{\rm m(k)}(\kv) |\psi (\kv)|^2 ,
 \eeq
where $\omega_i$ is an energy offset at the bottom of the polariton dispersion, i.e. $ \omega_{\rm i} =  \omega_{_{\rm L,U}} ({\bf 0})$, and $\psi (\kv)$ is the condensate wave function in $\kv$-space. It is valid, if and only if the main contribution stems from the second derivative of the dispersion relation. Here we have defined the kinetic energy dispersion $ \hat q^{\rm m(k)}(\kv)$, which corresponds to the dashed lines in Fig. \ref{F1}. In case this variation of the effective mass is further neglected we arrive at the simplest parabolic form for the kinetic energy \cite{ keen, keeli, Wout}, i.e.
\begin{equation}\label{par}
E^{\rm Par}_{\rm kin} = \int   \bigg( \frac{|\kv|^2}{2 m  ({\bf 0})} + \omega_{\rm i} \bigg) |\psi (\kv)|^2 =  \int  \frac{1}{2 m  ({\bf 0})}| \nabla \psi (\rv)|^2 +   \omega_{\rm i}  \int  |\psi (\rv)|^2,
\end{equation}
which corresponds to cGPE theory \cite{Light} and here the kinetic energy density is the dotted line in Fig. \ref{F1}. Next we address the question how these approximations are connected with the general dispersion relation. For the condensate wave function $\psi (\kv)$ in $k$-space it is given by 
\beq\label{haha2}
E^{}_{\rm kin} = \int \omega_{_{\rm L,U}} (|\kv|) |\psi (\kv)|^2.
\eeq 
The full polariton dispersion corresponds to the continuous line in Fig. \ref{F1} associated with the upper and lower branch respectively.
Furthermore by applying Taylor's theorem to $\omega_{_{\rm L,U}} (\kv)$ at $k = a =: |{\bf a}|$ we obtain
\begin{equation}\label{kg2}
E^{}_{\rm kin} = \int \Bigg( \omega_{_{\rm L,U}} (a)  + \omega'_{_{\rm L,U}} (a) (k-a) + \Bigg(\frac{\omega''_{_{\rm L,U}} (a)}{2} (k-a)^2 + \int^k_{a} (k-t)^2 \frac{\omega'''_{_{\rm L,U}} (t)}{2}   \Bigg) \Bigg) |\psi (\kv)|^2,
\end{equation}
which provides the context for the approximations discussed. Here the term $\omega'_{_{\rm L,U}} (a) (k-a)$ corresponds to the inertial mass which determines the wavepacket velocity from de Broglie's relation introduced in \cite{mk2}. It is zero when expanding at $ a = 0$. By dropping the remainder term in \eqref{kg2} we obtain the parabolic approximation \eqref{par}. In addition by Taylor's theorem we find that the $k$ dependent effective mass becomes a reasonable concept when expanding the dispersion for $k$ close to zero where $\omega'_{_{\rm L,U}} (0) = 0$ and when $\int^k_{0} (k-t) \omega''_{_{\rm L,U}} (t) \simeq k^2 \omega^{\prime\prime}_{\rm L,U}(k)$. 

In Fig. \ref{F1} we show the deviation between the approximate parabolic kinetic energy density, the kinetic energy density with velocity dependent effective mass and the complete kinetic energy density in terms of their $k$ dependence, indicating their range of validity. Fig. \ref{F1} (i) shows that the effective mass introduced in \cite{mk2} approximates the kinetic energy density from below and closer than the (parabolic) cGPE for the lower polariton energy branch. It overrates the kinetic energy shift for the upper branch as seen in (ii) with crossing of the curves at $k \simeq 7.5 \mu m^{-1}$ and it resembles the full dispersion curve more closely as compared with the parabolic dispersion approximation. However as the full dispersion and the effective mass concept allow larger $k$'s to be occupied at lower energy in the lower branch widening of the wave packet in $k$ space is expected and thus, due to the Fourier transform's duality properties, highly localised features in the wave packet are expected to emerge. We now look at how these different kinetic energy density approximations affect the resultant condensate model.

\newpage


\subsection{Spin sensitive state equations}  The kinetic energy term of a polariton condensate guiding equation may be defined in terms of a Fourier multiplier outside the Fourier transform of the condensate wavefunction \cite{mk}, i.e. in the form 
\beq\label{haha}
 \mathcal F^{-1} ( \omega_{_{\rm L,U}}(\kv) \mathcal F (f)) \equiv (q_{\rm L,U} \star  f) (\rv,t),
 \eeq 
 where $\omega_{_{\rm L,U}}(\kv) $ is real-valued and is the Fourier transform of $q_{\rm L,U}$ iff the transform exists and $\star$ denotes a convolution between two functions. Eq. (\ref{haha}) corresponds to a kinetic energy \eqref{haha2}. We consider this form of the kinetic energy for the spinor polariton field $\boldsymbol{\psi}$ componentwise, which is otherwise governed by cGPEs. These cGPEs are often coupled to a rate equation for the excitonic reservoir $n_R$ \cite{mumu, setup,Pinsker11}. Thus the two components' coupled wave equations are given by
\begin{widetext}
\begin{multline}\label{natural1}
i \hbar \partial_t  \psi_\pm (\rv,t) =   (1- i \eta) \cdot (q_{\rm L,U} \star  \psi_\pm) (\rv,t) + \\ + (1- i \eta) \bigg(  \alpha_1 ( |\psi_\pm|^2 + n^\pm_R) + V_{\rm ext} (\rv, t) + \alpha_2 |\psi_\mp|^2 \bigg) \psi_\pm (\rv,t)  +   \bigg( \frac{i }{2} \left(\gamma_C  n^\pm_R - \Gamma_d  \right) \bigg) \psi_\pm (\rv,t) .
\end{multline}
\end{widetext}
In \eqref{natural1} we consider a number of physical parameters: $\alpha_1 >0$ is the repulsive self-interaction strength, $\alpha_2 = -0.1 \alpha_1$ the attractive cross-interaction strength, $\gamma_C$ gives the scattering rate of the reservoir into the condensate, $\Gamma_d$ the decay rate of condensed polaritons and $\eta$ approximates the additional energy relaxation processes \cite{alex}. In general we also have an external time dependent potential $V_{\rm ext} (\rv, t)$. The reservoir dynamics are usually given by the rate equation \cite{Light} 
\beq
\partial_t n^\pm_R   = G^\pm_{\rm pump} -  n^\pm_R (\Gamma_R + \gamma_C ( |\psi_\pm|^2+ |\psi_\mp|^2)).
\eeq 
Here $G^\pm_{\rm pump} =G^\pm_{\rm pump}  (\rv,t) $ denotes the pumping distribution associated with the incoherent scattering into the polariton BEC of component $\pm$, and $\Gamma_R$ denotes the reservoir decay rate.  We  note that a formally equivalent equation has been suggested earlier for the atom laser based on atomic BEC \cite{AtomLaser}. For fast reservoir relaxation $\Gamma_R \gg \Gamma_d$ the reservoir dynamics are much faster than that of the condensate and the reservoir population can be approximated to leading order as   $n^\pm_R \simeq \frac{G^\pm_{\rm pump}}{(\Gamma_R + \gamma_C ( |\psi_\pm|^2+ |\psi_\mp|^2))}$ \cite{ alex, Pinsker11},
which for small amplitudes $ |\psi_\pm|^2$ can be further simplified. Consequently the growth and decay terms in \eqref{natural1} can be written in the  form \cite{Light, Pinsker11} so that 
\beq\label{approx}
i \hbar \partial_t  \rho_\pm |_{\rm gain/loss} =    i \left(P^\pm - \Gamma_\pm ( \rho_\pm+ \rho_\mp) - \gamma \right)  \rho_\pm
\eeq 
with $2 P^\pm =  \gamma_C G^\pm_{\rm pump}/\Gamma_R $,  $2 \Gamma_\pm = - \gamma^2_C G^\pm_{\rm pump}/\Gamma_R^2$ and $\gamma = \Gamma_d/2$ identified accordingly (while neglecting the relaxation contribution to the density occupation). Now to compare \eqref{natural1} with the corresponding effective mass approximations we generally write
\beq
 \mathcal F^{-1} ( \hat q (\kv) \mathcal F (f)) \equiv (q \star  f) (\rv,t),
 \eeq 
where $\hat q (|\kv|)$ corresponds to one of the three cases: $\hat q (k)=\omega_L(k)$, $\hat q (k)=\frac{k^2}{2}\partial_k^2\omega_L(k)+\omega_L(0)$ and $\hat q (k)=\frac{k^2}{2}\partial_k^2\omega_L(0)+\omega_L(0)$ as pointed out in the following discussions. 

\subsubsection{Unpolarised condensate}
When the condensate is unpolarised one considers a simple single component PDE to govern the condensate wave function:
\begin{equation}\label{singlecomp}
i \hbar \partial_t  \psi =   q \star  \psi + \bigg(\eta_1 n_R + \eta_2 P + \alpha |\psi|^2 + V(x)\bigg) \psi -   i \gamma \psi + i \kappa n_R \psi,
\end{equation}
with $\Gamma_d/2 = \gamma$ and where 
\beq\label{res2}
n_R \simeq \frac{G_{\rm pump}}{\Gamma_R + \gamma_C |\psi|^2},
\eeq
which is the spin coherent counterpart of $n_R^\pm$ and its approximation physically justified on the same grounds.
One can further simplifying the reservoir and decay of polaritons via a Taylor series given by \cite{Pinsker11}
\beq
\kappa n_R \simeq \kappa G_{\rm pump}/\Gamma_R (1- \gamma_C/\Gamma_R  |\psi|^2) \equiv \left( P - \Gamma |\psi|^2  \right).
\eeq
In addition, for small amplitudes of the wave function, one may approximate $\eta_1 n_R + \eta_2 P  \simeq \beta n_R$.

\subsection{Unpolarised plane waves} 
\begin{figure}[ht]
\vspace{20mm}
\begin{tabular}{c}
\begin{picture}(150,0)
\put(-220,-90) {\includegraphics[scale=0.21]{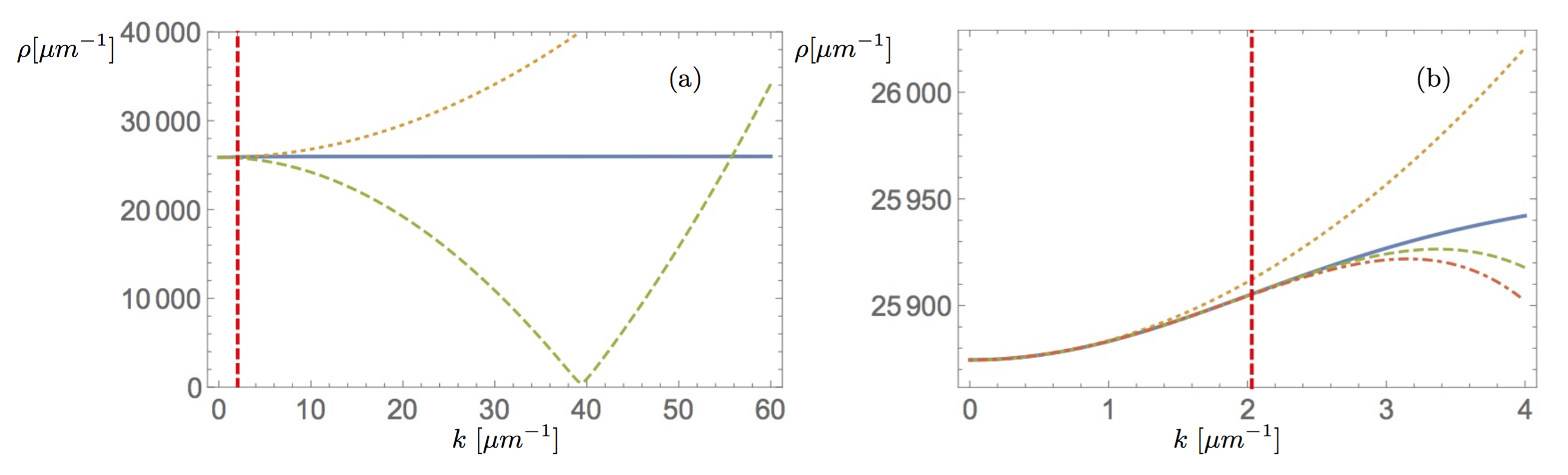} }
\end{picture} 
\end{tabular}

\vspace{30mm}

\caption{ (a) The density of plane wave solutions for the lower polariton branch: Solid line corresponds to the plane wave for the lower polariton dispersion, dotted line to the parabolic approximation with energy shift and the dashed line to the plane waves according to an effective negative mass model. The vertical red-dashed line is at the inflection point. (b) Plane wave solutions of the lower branch: The lines are associated as follows: Continuous line corresponds to the full dispersion, the dotted line to the parabolic dispersion with offset, the dashed line to the Taylor approximation of the dispersion at $k=0$ to order four and the dotted-dashed line to the order six.}
\label{jo}
\end{figure}

After discussing the explicit guiding equations we now turn to results and predictions which deviate from the simpler approximations to the polariton kinetic energy \eqref{haha2}.  We begin by considering the stationary (i.e. time-independent) plane wave solutions of polariton condensates in the presence of pumping and decay processes, i.e. solutions of the equation
\beq\label{hoho}
 q \star \psi + \alpha_1 |\psi|^2  \psi + i \left( P - \Gamma_d |\psi|^2  - \gamma_c \right) \psi = (\mu-\beta n_R) \psi.
\eeq
To solve this model we  take the stationary ansatz $\psi = v  \exp(i \kv_{\rm i}\cdot \rv)$ such that 
\beq
 q \star \psi = q\star(v  \exp(i \kv_{\rm i} \rv))  = v \cdot \mathcal F^{-1}( \omega(\kv)  (2\pi) \delta(\kv - \kv_{\rm i}) )= \omega(\kv_{\rm i}) v  \exp(i \kv_{\rm i} \cdot \rv) = \omega(\kv_{\rm i})\psi.
 \eeq
Consequently we get an algebraic equation for the plane wave amplitudes
\beq
\mu - \beta n_R= \left(\omega (\kv_{\rm i}) + \alpha_1 |v|^2 + i \left( P - \Gamma_d |v|^2  - \gamma_c \right) \right),
\eeq
which has complex solution
\beq
v(\kv_{\rm i}) = \pm \sqrt{\frac{i \gamma_c - \mu - \beta n_R - i P_1 - \omega(\kv_{\rm i})}{\alpha_1 - i \Gamma_d}}.
\eeq
The complete solution is therefore given by
\beq
\psi (\rv )=  \exp{(i c)} \sqrt{\frac{i \gamma_c - \mu - \beta  n_R - i P_1 - \omega(\kv_{\rm i})}{\alpha_1 - i \Gamma_d}}\exp(i \kv_{\rm i}\cdot \rv)
\eeq
for any $c \in \mathbb R$. In Fig. \ref{jo} (a) and (b) we present the $1$d plane wave density $\rho (k_{\rm i}) = |\psi (\rv, \kv_{\rm i})|^2$ as a function of the dispersion relation indicating the inherent differences of the kinetic energy models and the actual/observable plane waves. While approximations yield results/plane waves that are similar to the full dispersion model when $k$ is below the inflection point the large $k$ behaviour differs significantly.  

\subsubsection{Superposition in periodic potential and the equilibrium no-interaction case}

As an example for deriving dynamical equilibrium plane waves we consider the generalised dispersive PDE for a wave moving in the reference frame with velocity $\vv$ in a periodic potential $V_{\rm per}$, i.e.
\beq
 q \star \psi + \alpha_1 |\psi|^2  \psi -  V_{\rm per} \psi +i \vv \nabla \psi = i \hbar \partial_t \psi.
\eeq
Now let us make the separation of variables ansatz for a superposition of two different k-modes as a toy model of a wave packet and to study the dispersion between the two $k-$modes:
\beq
\psi = p(t)  \exp(i \kv_{\rm i}\cdot \rv) + g(t) \exp(i \kv_{\rm b}\cdot \rv)
\eeq
First we note the linearity of the $q \star$ operator, i.e.
\begin{multline}
 q \star \psi  = p(t) \cdot \mathcal F^{-1} \left( \omega(\kv)  (2\pi) \delta(\kv - \kv_{\rm i}) \right) + g(t) \cdot \mathcal F^{-1} \left( \omega(\kv)  (2\pi) \delta(\kv - \kv_{\rm b}) \right) = \\ = \omega(\kv_{\rm i}) p(t)  \exp(i \kv_{\rm i}\cdot \rv) + \omega(\kv_{\rm b}) g(t)  \exp(i \kv_{\rm b}\cdot \rv).
\end{multline}
On the other hand we have to satisfy
\beq
i \hbar \partial_t p = (\omega (\kv_{\rm i} ) + \alpha_1 |p|^2 - \kv_{\rm i}\cdot \vv ) p
\eeq
and
\beq
i \hbar \partial_t g = (\omega (\kv_{\rm b} ) + \alpha_1 |g|^2 - \kv_{\rm b}\cdot \vv ) g,
\eeq
if the external potential is given by $V_{\rm per} = - 2 \alpha_1  {\mathcal Re} (p g) \cos((\kv_{\rm i}-\kv_{\rm b}) \rv)$. Thus we obtain the analytical condensate wave function
\begin{multline}\label{wavepacket}
\psi (\hbar t, \rv) = A \exp(-i \omega(\kv_{\rm i}) t) \exp(-i \alpha_1|A|^2 t) \exp(i \kv_{\rm i} \vv t) \exp(i \kv_{\rm i} \rv) + \\ + B \exp(-i \omega(\kv_{\rm b}) t) \exp(-i \alpha_1|B|^2 t) \exp(i \kv_{\rm b} \vv t) \exp(i \kv_{\rm b} \rv).
\end{multline}
Assume $a>b$ then it implies for the monotonically increasing polariton dispersion that $\omega(a) > \omega (b)$. If we assume that the interference of the nonlinearity is vanishing and there is no external potential we would observe the same behaviour. Here note that the ``wave packet" consisting of two plane waves is coherent, iff
\beq
 (-\omega(\kv_{\rm b}) - \alpha_1|B|^2  + \kv_{\rm b} \vv)/| \kv_{\rm b}|  =  (-\omega(\kv_{\rm i}) - \alpha_1|A|^2  + \kv_{\rm i} \vv )/| \kv_{\rm i}| . 
\eeq
Any localised structure, such as moving bright solitons in BEC, has to satisfy such a condition component-wise, which is in particular satisfied for certain wave-packets in the focusing case $\alpha_1 <0$.  Here the attractive interactions cancel out the dispersive effects from the kinetic energy due to opposing sign \cite{Tao}. Note that the theoretical description in the inertial frame moving with $\vv$ the plane waves does not include a Doppler term $i \vv \nabla \psi $ and thus in such a frame the condition
\beq
 (-\omega(\kv'_{\rm b}) - \alpha_1|B'|^2)/|\kv'_{\rm b}|   =  (-\omega(\kv'_{\rm i}) - \alpha_1|A'|^2)/|\kv'_{\rm i}|
\eeq
applies. In this sense temporal coherence of a wave packet could be feasible, however note that taking into account interference terms due to the nonlinearity will modify this behaviour.

\subsection{Spin sensitive results for plane and linear waves}

The simplest scenario to begin an examination of the effects of the dispersion relation in spin sensitive systems is the case of a homogeneous external potential (absorbed in the chemical potential), with pumping and decay in the simplest approximate form \cite{Pinsker11}, where the incoherent mode equation  becomes
\begin{equation}\label{natural2}
  q \star \psi_\pm + (\alpha_2 |\psi_\mp|^2+\mu) \psi_\pm    + (\alpha_1 |\psi_\pm|^2 + n_R) \psi_\pm =  i \frac{ \left( P
  - \Gamma (|\psi_\pm|^2 + |\psi_\mp|^2) - \gamma \right)}{(1- i \eta)} \psi_0 
\end{equation}
and includes energy relaxation processes via $(1- i \eta)$. We simplify the consideration to $(1+1)$d and note that the higher dimensional case is analogous. Furthermore we use the notation $\omega(a) = \hat q(a)$ to relate the dispersion to the $x$-space operator $q(x)$.
The ansatz for a stationary solution is $\psi_\pm (x,t) = \phi^\pm_0 \exp(i a x)  \exp(i \mu' t)$ with $\mu' = \mu (1- i \eta)$ and by recognising the translation property of the Fourier transform
$\mathcal F \left( \psi_\pm \right) = \phi^\pm_0  \delta (k - a)$ 
 we write \eqref{natural2} as
\begin{equation}
(1 - i \eta)\bigg[\hat q(a) + \mu  +
\alpha_1 (|\phi^\pm_0|^2 + n_R  ) + \alpha_2 |\phi^\mp_0|^2 \bigg]   =   i\left(P_\pm  -  \Gamma_\pm (|\phi^\pm_0|^2 + |\phi^\pm_0|^2 ) - \gamma \right).
\end{equation}

We solve the spin sensitive system under the simplifying (but not necessary) assumption $\Gamma_\pm = \Gamma$ by the analytic plane wave solutions  for the two polariton spin components $\pm$,
\begin{widetext}
\begin{equation}\label{homot2}
( \phi^\pm_0 )^2=  \frac{ \Gamma \Delta -\alpha_1^2(i + \eta)^2 n_R + \alpha_2(i + \eta)( (i+\eta) (\mu + \hat q) + P_\mp -\gamma) - \alpha_1 (i + \eta)( (i+\eta) ( \hat q + \mu-\alpha_2  n_R ) + P_\pm -\gamma)}{(\alpha_1 - \alpha_2) (i + \eta) ((\alpha_1 + \alpha_2)(i + \eta)-2 \Gamma)},
\end{equation}
\end{widetext}
introducing the pump detuning between components $\Delta = P_\pm - P_\mp$. Here the parameter $a$ defines the position on the polariton dispersion branch altering the wave formation via the kinetic energy density $\hat q=\hat q(a)$ in Fig. \ref{F1}.  Generally we have $\hat q(k) = \omega_{_{\rm L,U}} (k)$ while in the approximated velocity dependent mass case $\hat q(k) =\frac{k^2}{m_{\rm L,U} (k)}$, which by setting $m_{\rm L,U} (k) \to m_{\rm L,U} (0) = const.$ resembles the parabolic case. The presence of the reservoir decreases the plane wave amplitude while the opposite spin component increases the amplitude and vice versa consistent with the analysis in \cite{Pinsker11}. We note that for slowly varying $\psi_\pm (x)$ and $P_\pm(x)$ we set $P_\pm  \to P_\pm(x)$ in \eqref{homot2}.

 \begin{figure*}[]
 \vspace{20mm}
 \begin{tabular}{c}
\vspace{41mm}
\begin{picture}(150,-55)
\put(-194,-120) {\includegraphics[scale=0.2]{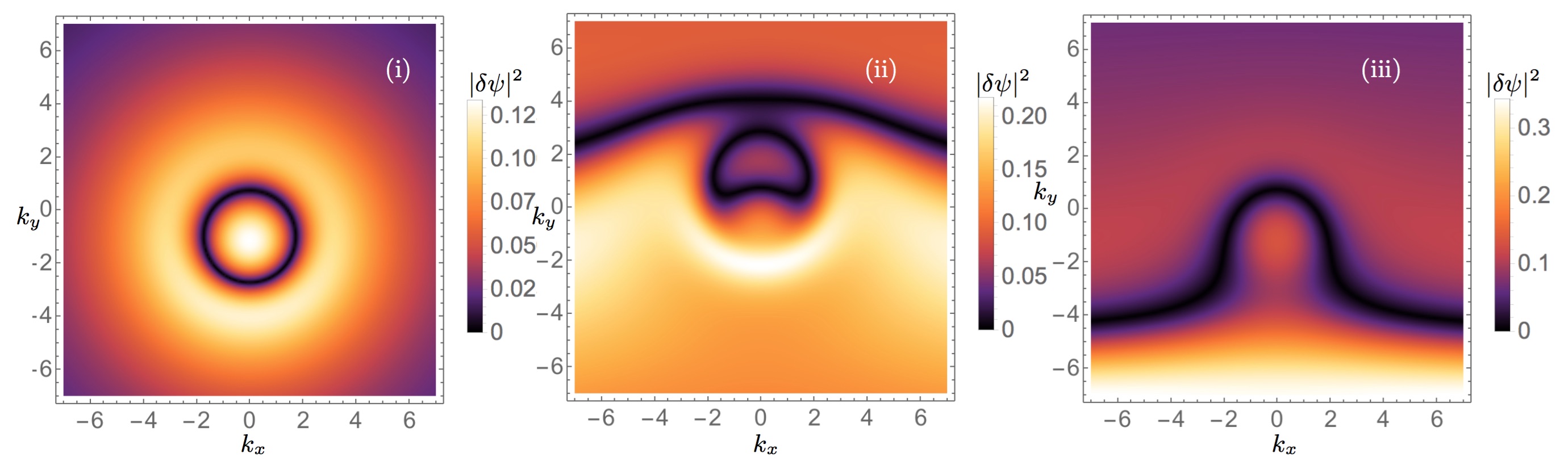} }
\end{picture} \hspace{2mm} 

\end{tabular}
\vspace{0mm}

\caption{Linear wave functions $|\dpsi(k_x,k_y)|$  of the lower polariton branch for the parabolic (i), the $m(k)$ model (ii) and the general  kinetic energy (iii) showing the structural differences of the predictions of the three kinetic energy models. The direction of motion is in $-k_y$ direction and the specific parameters corresponding to the figures are set as follows. $b \tilde P^* = 1$, $P=1$, $v_x = 0$, $v_y =-1$, ${\mathcal Im} (a+b)^2 +{\mathcal Re} (a)^2 +{\mathcal Re} (c)^2 + 2 {\mathcal Re} (a+c) + 2 a {\mathcal Re} (c)  =13$, ${\mathcal Re} (a+c) =1$. Further we subtract $\omega_L(0)$ from the dispersions for better visibility.}
\label{FL}
\end{figure*}
It is useful to separate the exact motion into bulk motion plus low amplitude acoustic disturbances to understand the elementary excitations \cite{Pupu}. So we consider the linear waves by making an ansatz of the form $
\psi_\pm(\rv,t) 
\equiv \phi_\pm e^{-i \mu t}  + \dpsi_\pm (\rv) e^{- i \mu t} + \dpsi^*_\pm (\rv) e^{-i \mu t}$,
where $\phi_\pm$ represents the unperturbed part solving the mean-field model such as the plane wave solutions presented above and the linear waves $\dpsi (\rv)= (2\pi)^{-D/2} \int \dpsi_k \exp{(i \rv \cdot \kv)}$. By inserting in the spin sensitive PDEs and dropping terms of order $\dpsi^2$ and including a chemical potential $\mu_\pm = n_\pm \alpha_1$, we get the Bogoliubov equations in $\kv$-space for the linearised perturbation dynamics, i.e.
\begin{widetext}
\begin{multline}\label{lini1x3}
 i \frac{\partial \dpsi_\pm}{\partial t} = q \star \dpsi_\pm+ \alpha_1 \left(2 |\phi_\pm|^2 - \frac{\mu}{\alpha_1} \right) \dpsi_\pm + \alpha_1 \phi_\pm^2 \dpsi^*_\pm + \alpha_2 ( |\phi_\mp|^2 \dpsi_\pm + \phi_\mp \phi_\pm \dpsi^*_\pm + \phi^*_\mp \phi_\pm \dpsi_\mp) \\ -i 2 \Gamma |\phi_\pm|^2 \dpsi_\pm  + i (P_\pm (\rv) - \gamma) \dpsi_\pm - i \Gamma \phi_\pm^2 \dpsi^*_\pm - i \Gamma ( |\phi_\mp|^2 \dpsi_\pm + \phi_\mp \phi_\pm \dpsi^*_\pm + \phi^*_\mp \phi_\pm \dpsi_\mp ).
\end{multline}
To arrive at those equations we have used the linearizations $ |\psi_\pm|^2 \psi_\pm \simeq 2 |\phi_\pm|^2 \dpsi_\pm + \phi^2_\pm \dpsi^*_\pm+ |\phi_\pm|^2 \phi_\pm$ and $ |\psi_\pm|^2 \psi_\mp \simeq |\phi_\pm|^2 \dpsi_\mp + \phi_\pm \phi_\mp \dpsi^*_\mp + \phi^*_\pm \phi_\mp \dpsi_\pm +  |\phi_\pm|^2 \phi_\mp$, i.e. dropping all terms of order $\dpsi^2$ and higher, $|\phi_\pm|^2 \phi_\pm$ and $ |\phi_\pm|^2 \phi_\pm$  for the excitations dynamics, i.e. we have separated those equations from the bulk dynamics.  So the bulk of the spinor condensate is guided by 
\begin{equation}\label{model1x1}
 i \frac{\partial \phi_\pm}{\partial t} = q \star \phi_\pm +  (\alpha_1 |\phi_\pm|^2 + \alpha_2 |\phi_\mp|^2) \phi_\pm + i (P_\pm (\rv) -  \Gamma (|\phi_\pm|^2 + |\phi_\mp|^2 ) - \gamma) \phi_\pm.
\end{equation}
Assuming the pumping function $\tilde P_\pm = (2\pi)^{-D/2} i \int e^{-i \kv \rv } (P_\pm- \gamma)\dpsi_\pm$, and by introducing the abbreviations $a = \alpha_1 ( |\phi_\pm|^2 - \frac{i 2 \Gamma_\pm}{\alpha_1} |\phi_\pm|^2 ) + (\alpha_2 - i \Gamma_\pm) |\phi_\mp|^2$, $b=\alpha_1 \phi_\pm^2 - \frac{i 2 \Gamma_\pm}{\alpha_1} |\phi_\pm|^2 +(\alpha_2 - i \Gamma_\pm)\phi_\mp \phi_\pm$ and $c = (\alpha_2 - i \Gamma_\pm)  \phi^*_\mp \phi_\pm $ we exactly solve Eq. (\ref{lini1x3}) for $ \phi_\mp = \phi_\pm$, and $\tilde P_- = \tilde P_+$. The result is
\beq\label{define2}
 \dpsi^\pm_k = \frac{b \tilde P^*_\pm - \tilde P_\pm \left( \hat q (k) + a^* + c ^* - \kv \vv \right)}{\hat q^2(k) - |b|^2 + {\it Im} (a + c)^2 + {\it Re} (a)^2 + {\it Re} (c)^2 + \kv \vv^2 + 2 {\it Re} [ (a+c-\kv \vv) \hat q(k) + a  {\it Re} (c) -\kv \vv  {\it Re} (a+c)] }.
\eeq
\end{widetext}
Here for a given and implicitly defined $\tilde P_\pm$ we obtain the excitations $ \dpsi^\pm_k$, which in turn defines implicitly the physical pumping function $P_\pm$.
Furthermore for ${\it Re} (c) \to 0$ and ${\it Im} (c) \to 0$ we recover the spin coherent case. We observe a significant modification of the polariton excitation formation by considering the wave packets explicitly given by \eqref{define2} due to the functional variation of $\hat q(k)$ as presented in Fig. \ref{FL}. Here the direction of propagation of the linear waves is downwards ($v_x = 0$ and $v_y = -1$ with $\vv =(v_x ,v_y )$) implying symmetry breaking of the ring structure in the general framework thus showing severe changes in the phenomenology of the polariton linear waves as compared with cGPE-type excitations. More specifically, in Fig. \ref{FL} (i) we observe a ring structure within the linear wave solution $|\dpsi^\pm (k_x ,k_y )|$ when a parabolic approximation is assumed. However, the more general model $\hat q (k)=\frac{k^2}{2}\partial_k^2\omega_L(k)+\omega_L(0)$ corresponding to Fig. \ref{FL} (ii) shows linear waves that break this ring symmetry, particularly by including a singular line orthogonal to the direction of motion. Finally Fig. \ref{FL} (iii) represents the most accurate linear wave of the general dispersion, i.e.  $\hat q (k)=\omega_L(k)$, and again shows the breaking of the ring's symmetry by a disconnected singular line. This breaking of the ring symmetry is not observed when the linear wave is moving at lower velocities or at rest, as observed as well for the case of its parabolic approximation.

Following our presentation of  analytical wave results we now turn to the numerical phenomenology due to the different kinetic energy model predictions.

\subsection{Conservative localized pulse evolution}

\begin{figure}[]
 \vspace{20mm}
 \begin{tabular}{c}
\vspace{41mm}
\begin{picture}(150,-55)
\put(-227,-100) {\includegraphics[scale=0.23]{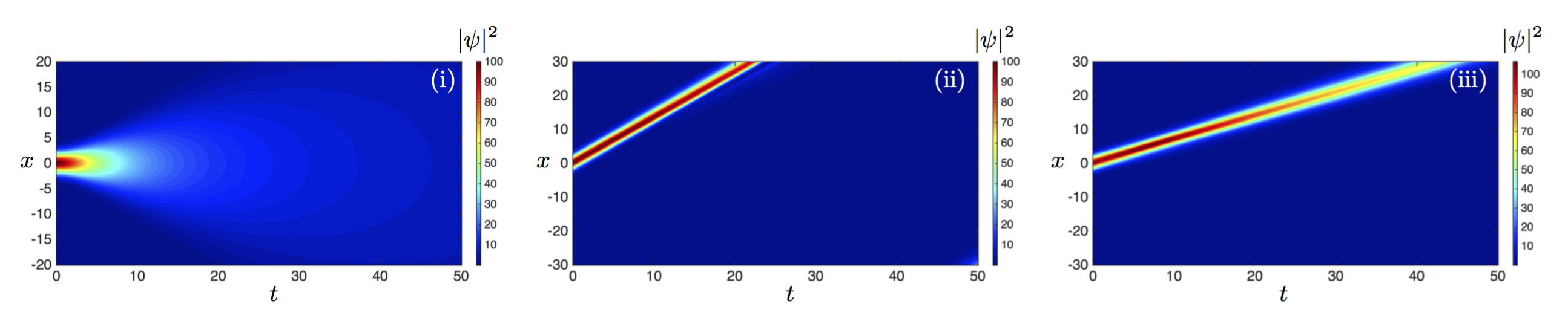} }
\end{picture} \hspace{0mm} 
\end{tabular}
\vspace{-10mm}
\caption{Density plots  $|\psi(x,t)|^2$ of the equilibrium condensate wave function predictions for the full polariton dispersion at different $k$. (i) corresponds to $k=0$, (ii) to $k=2.2$ and (ii) to $k =4$.  In the simulations a $4$th order time splitting in time with $\Delta h=0.005, \Delta t=0.005$ was used. A very large domain $[-50,50]$ was employed to make sure the value of the solution on the boundary is negligible.  Numerical parameters were  $\alpha = 0.001$, $V_0 = 0$ and $\delta = 0.004$.\label{F22}}
\end{figure}

\begin{figure}[]
 \vspace{10mm}
 \begin{tabular}{c}
\vspace{41mm}
\begin{picture}(150,-55)
\put(-227,-100) {\includegraphics[scale=0.275]{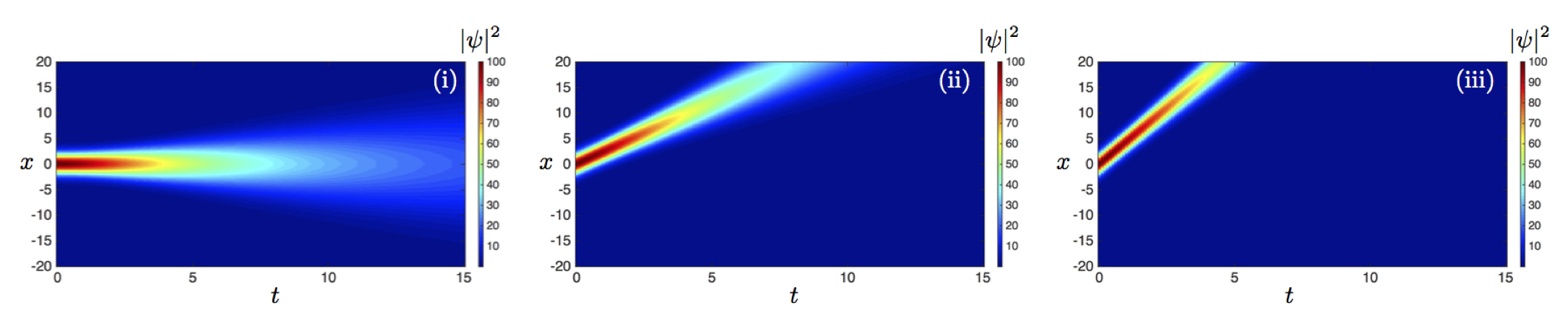} }
\end{picture} \hspace{0mm} 
\end{tabular}
\vspace{-5mm}
\caption{Density plots  $|\psi(x,t)|^2$ of the equilibrium condensate wave function predictions for the parabolic polariton dispersion at different $k$. (i) corresponds to $k=0$, (ii) to $k=2.2$ and (ii) to $k =4$.  In the simulations a $4$th order time splitting in time with $\Delta h=0.005, \Delta t=0.005$ was used. A very large domain $[-50,50]$ was employed to make sure the value of the solution on the boundary is negligible.  Numerical parameters were  $\alpha = 0.001$, $V_0 = 0$ and $\delta = 0.004$. \label{F222}}
\end{figure}

First we consider the scenario of a localised wave packet neglecting the non-equilibrium effects and assuming that the condensate evolves without an external potential. Thus the governing equation of motion is
\begin{equation}
i \hbar \partial_t \psi  =   q \star \psi  +  \big( \alpha |\psi|^2 + n'_R + V  \big) \psi,
\end{equation}
where  $V(x) = V_0(x) + \delta P(x)$ and $\kappa/\alpha = 1.36$. As an initial condition we first set 
\beq
\psi(x,0) = A_p \exp\left(-\frac{W x^2}{2} \right)  \exp(-i k x),
\eeq
while we choose the numerical parameters $A_p = 10$, $W = 0.25$ and $k=0,2.2,4$. In Fig. \ref{F22} we show that spatial localisation of the initial data can be preserved to a higher degree above the inflection point at $k \simeq 2 \mu m^{-1}$ despite the dispersive character of the kinetic energy. For $k=0$ the initial wave function disperses rapidly. Corresponding results for the parabolic kinetic energy show stronger dispersion for large $k$, i.e. above the inflection point as presented in Fig. \ref{F222}. These observations mimic the behaviour of bright-type solitons observed in \cite{negative}. We note also that the velocity of the wave packet reduces at larger $k > k_{\rm inf}$, when considering the full dispersion relation, as the velocity of the wavepacket is given by the derivative of the dispersion. At $k=4$ the dispersion is flatter than at $k=2.2$, so the wavepacket is slower (compare the slope of the trajectory in Fig. \ref{F22} (b) and (c)). This stands in stark contrast to the velocity due to the parabolic dispersion, which increases with $k$ (compare Fig. \ref{F222} (b) and (c)).

These results are in agreement with the observation by Sich {\em et al.} \cite{negative} of localisation occuring above the inflection point.  We note however one significant difference in the modeling.  In \cite{negative} the excitonic and photonic modes were explicitly modeled, with the complex dispersion profile of the polariton condensate emerging as a consequence of the coupling between the modes.  Here instead we model only the polariton mean-field, and explicitly include the dispersion relation in the dynamics, allowing us to identify directly the contribution of the dispersion on the resulting dynamics.  

\subsection{Dark soliton instability} 
\begin{figure}[]
 \vspace{20mm}
 \begin{tabular}{c}
\vspace{41mm}
\begin{picture}(150,-55)
\put(-210,-100) {\includegraphics[scale=0.257]{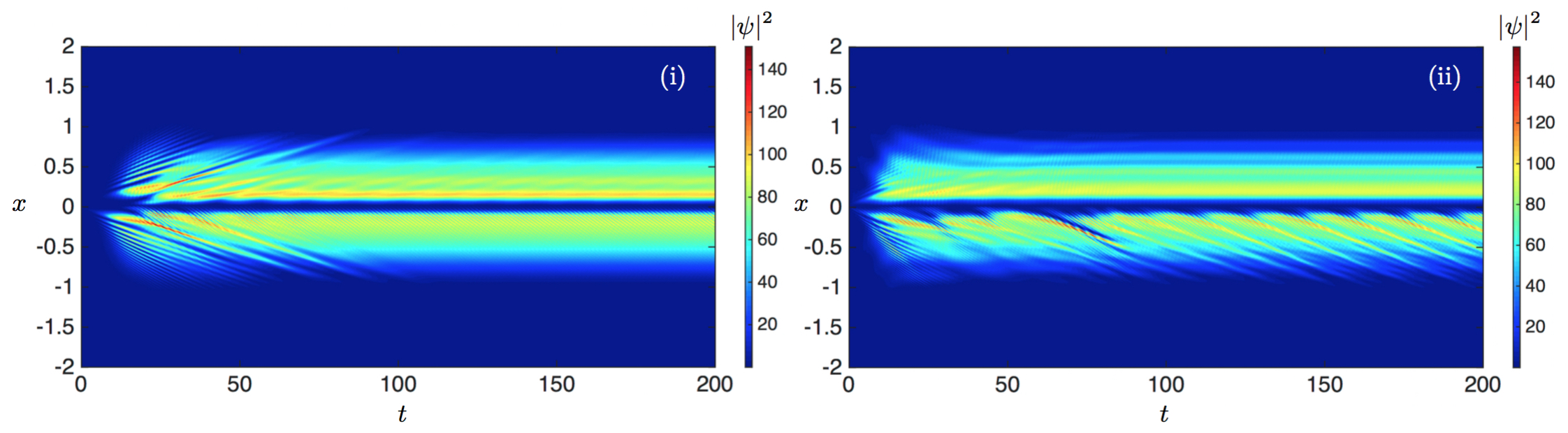} }
\end{picture} \vspace{-5mm} 
\end{tabular}
\caption{Density plots $|\psi(x,t)|^2$ of the condensate wave function predictions for the case dynamics with $k$ sensitive mass (i) and for the full lower polariton dispersion with a weak self-interaction strength. In the simulations a $4$th order time splitting in time with $\Delta h=0.005, \Delta t=0.005$ was used. A very large domain $[-50,50]$ was employed to make sure the value of the solution on the boundary is negligible.  Numerical parameters were  $\alpha = 0.001$, $V_0 = 5$, $A_P= 6.6$, $\sigma = 1$, $n'_R \simeq  P (x)/10 (1 - 0.005 |\psi|^2)$, $\delta = 0.004$ and $\gamma = 0.5$. \label{F2}}
\end{figure}

\begin{figure}[]
 \vspace{20mm}
 \begin{tabular}{c}
\vspace{41mm}
\begin{picture}(150,-55)
\put(-205,-140) {\includegraphics[scale=0.26]{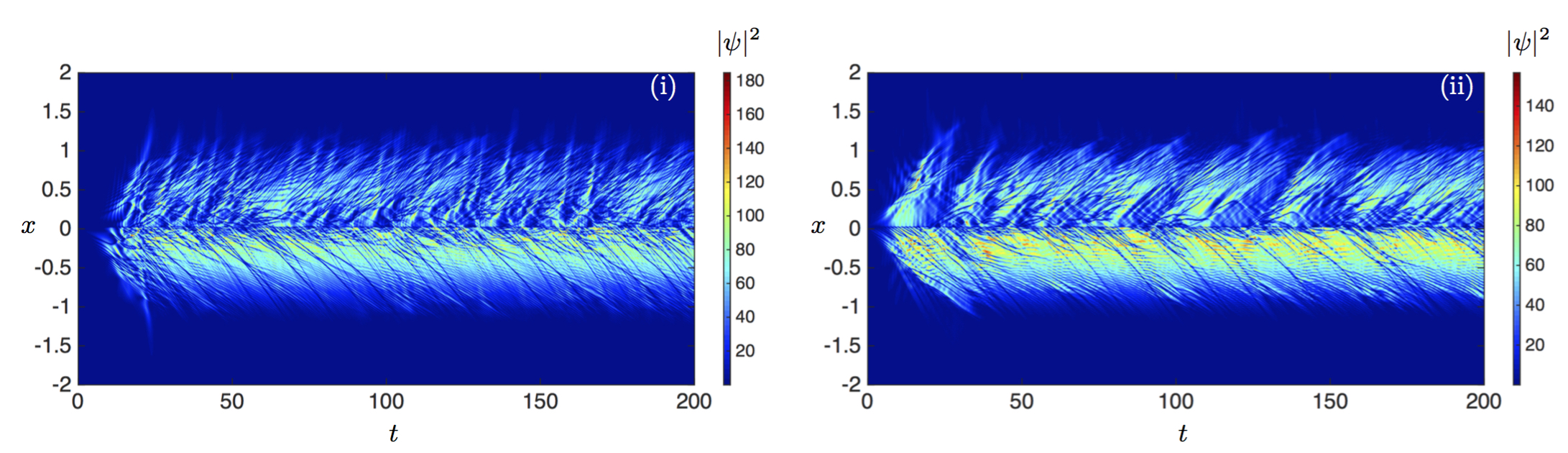} }
\end{picture} \vspace{5mm} 
\end{tabular}

\caption{Density plots $|\psi(x,t)|^2$ of the condensate wave function predictions for the kinetic energy of the effective mass model (i) and the general polariton dispersion (ii)  with a stronger self-interaction strength which is $50$ times larger than the one used for Fig. \ref{F2}. }
\label{F3}
\end{figure}

Following from \eqref{natural1} and \eqref{approx} the single polariton state equation resembling an incoherent driving scheme can be approximately written as
\begin{equation}
i \hbar \partial_t \psi  =  (1- i \eta) \cdot (q \star \psi  +  \big( \alpha |\psi|^2 + V  \big) \psi)  +  i (\kappa \cdot n'_R - \gamma) \psi,
\end{equation}
where $n_R' = \alpha n_R$, $V(x) = V_0(x) + \delta P(x)$ and $\kappa/\alpha = 1.36$. 
Utilising this model in \cite{setup} the generation of dark soliton trains in $D=1$ within an experimentally accessible scheme has been demonstrated, i.e. for wire shaped micro-cavities embedding a metallic decomposition on the half-line. As in  \cite{setup} we set the potential in $x$-space due to this metallic contact to be 
\beq
V_0 (x) = 
\begin{cases}
 V_0 > 0  \hspace{6mm}\text{ for    } x \geq 0 \\  V_0 =  0 \hspace{6mm}\text{ for    }  x < 0 
\end{cases}
\eeq
and assume a Gaussian pumping spot resembling the spatial form of the incoherent polariton ground state formation, 
\beq
P (x) = A_{\rm P} \exp{\left(- \frac{x^2}{\sigma^2} \right)}.
\eeq
We refer to \cite{PRL, PRX,Light} for various applications of this model, when we assume a parabolic dispersion.
As shown in Ref. \cite{dark} a local abrupt change of interaction strength of a condensate establishes a stable and regular dark soliton train within a conservative GP theory. Once the flow in the direction of decreasing interaction due to particle repulsions is locally crossing the speed of sound $c_s(x)=\sqrt{\mu(x)/m}$, where $\mu(x)=\alpha n(x)$ for a scalar condensate,  dark solitons are formed from dispersive shock waves at the point of abrupt change in self-interactions\cite{Kamchatnov}.  These solitons then proceed to dissipate the local excess of energy \cite{setup}. While in polariton condensates the interaction strength $\alpha_1$ can be varied by tuning the exciton/photon detuning, and there is an ongoing debate on its experimentally measured value \cite{large}, it is straightforward to apply a tunable potential step $V_0(x,t)$. The mechanism for soliton generation is again the breaking of the sound-barrier in the region $x<0$ in the presence of a perturbation at $x=0$ \cite{setup}. In the regime of soliton-train generation, the frequency $\nu$ increases with the magnitude of the potential step as the corresponding increase of mass passing the step at $x=0$ allows a more frequent breaking of the local speed of sound. In Fig. \ref{F2} and \ref{F3} we observe a strong modification of the mean-field density dynamics due to the non-parabolic dispersion relation, as compared with the regular dark soliton train patterns of cGP-theory reported in \cite{setup}. We find that at low $k$ the kinetic energy is in a quasi-parabolic regime supporting dark soliton solutions consistent with the graphs presented in Fig. \ref{F1}. We see also in Fig. \ref{F2}  that the effective mass induces additional bright-type high-density waves of varying frequency that fade out for larger times in (i) while they persist with a fixed frequency in the full dispersion regime (ii) on top of the regular dark soliton train arrays - a phenomenon entirely unobserved or neglected in cGP theory. We attribute this localisation phenomenon to the flat part of dispersion relation in the lower polariton branch.  Finally we note that larger self-interaction strengths, as reported in \cite{large}, would lead to chaotic dark soliton trains Fig. \ref{F3} (ii) while again stable patterns are reported for the parabolic dispersion approximation for appropriate parameters, thus providing an indirect test for the large interaction hypothesis.

\section{Discussion} The topic of kinetic energy and particularly negative effective mass of polaritons has received increasing attention \cite{mk, mk2,negative,negative2}. Here we find that the effects on the pattern formation can be interpreted in the way suggested so far only for small $k$ below the inflection point, because the kinetic energy of the polariton is composed of an array of terms (which can be stated in terms of Taylor's theorem). To directly and unambiguously observe the switch in sign of the effective mass, i.e. the second derivative term plus offset within the condensate wave function, as e.g. reported in \cite{negative}, would require a situation including physical processes that neutralise all the other terms in the expansion \eqref{kg2}. Energy shifts are obtained by external potentials, however the variation of the remainder terms at the inflection point $k \sim 2 \mu m^{-1}$ of the lower polarition dispersion is larger than that of the effective mass and thus overlaps its sign switch - the dispersion is positive for all $k$. Furthermore the dispersions of the polariton branches are monotonically increasing in $k$ and thus there is no switch in sign of the  kinetic energy. This implies that e.g. the time dependent phase of plane wave solutions due to the kinetic energy cannot be cancelled out by the contribution due to nonlinear interactions.  This in turn puts into question the mechanism of bright solitons reported in \cite{negative}, since those bright solitons are agued by the observation that negative mass and repulsive self-interactions are formally equivalent to positive mass with attractive interactions, i.e. $i \partial_t \psi(x,t) =\mu \psi(x) = \big( \mp \Delta \mp |\psi|^2 \big) \psi(x)$, with chemical potential $\mu = \mp n(0)/2$. Such a model implies the well-know bright soliton solutions, because the dispersive effects are compensated by the attractive self-interaction forces. Now when considering the toy model case of a superposition of two plane waves including the polariton dispersion as kinetic energy we have found that they move coherently in time, iff
\beq
 (-\omega(\kv'_{\rm b}) - \alpha_1|B'|^2)/|\kv'_{\rm b}|   =  (-\omega(\kv'_{\rm i}) - \alpha_1|A'|^2)/|\kv'_{\rm i}|,
\eeq
when interference terms are neglected. This illustrates that due to $\alpha_1, \omega >0$ only for specific amplitudes coherent motion is feasible. Such condition could be satisfied in \cite{negative}. Furthermore we have simulated the movement of a gaussian wave packet at different positions of the dispersion and can indeed confirm that the dispersion of the wave packet is reduced above the point of inflection, when considering the full polariton kinetic energy - the spatial localisation of an initial wave is stronger in the more accurate full dispersion model compared to predictions of the parabolic cGP, thus supporting the observation of localisation or suppressed dispersion in \cite{negative}. However note that the term soliton or bright soliton is defined rigorously \cite{Book}  and involves properties such as unchanged shape and speed over time and particularly after a collision with another soliton and thus one may more accurately refer to these temporarily localised structures as near-bright-type solitons.

We suggest that localisation of a wave packet in $x$-space is due the flat dispersion relation of the lower polariton branch allowing large $k$ modes to be occupied with less energy than in cGPE models. In addition we note that an alternative route to bright soliton generation in polariton condensates could be along the lines of effective attractive self-interactions as discussed in \cite{bright2} with results similar to matter-wave bright soliton formation in ultra-cold lithium-$7$ gases \cite{real}.  Apart from this the signs of the non-parabolic dispersion relation of polaritons are apparent in many aspects of wave formation starting from bright-type solitons trains on top of dark solitons trains to chaotic dark solitons to the explicit form of linear waves of the polariton spin modes. Thus for accurate description of the polariton mean field mode the full dispersion should be included in future dynamical models for more accurate predictions.

\section{Methods}

We used a generalised Gross-Pitaevskii theory to analyse the impact a general form for the kinetic energy has on possible polariton condensate dynamics. The mathematical analysis presented is based on standard analytical tools, complex algebra and integration, and the Bogoliubov approach to linear waves. An efficient numerical method was employed for the simulation of the partial differential equation. In time, the second order time splitting method, i.e. the Strang splitting, is used for the time evolution. In space, the wave function $\psi(x,t)$ is discretized on uniform grids and the spectral method based on the Fourier series is used to deal with the generalized kinetic energy term $q*\psi(x,t)$. 
 For one time step, the algorithm works as follows:
\begin{itemize}
\item solve $i\partial_t\psi(x,t)=q*\psi(x,t)$ for half a time step via the discrete Fourier-Transform (DFT).
\item solve the remaining part for one time step via the 4th order Runge-Kutta method.
\item solve $i\partial_t\psi(x,t)=q*\psi(x,t)$ for another half time step.
\end{itemize}
In the simulations, we choose $\Delta h=0.005$ and $\Delta t=0.005$. A large domain $[-50,50]$ is used to make sure the solution on the boundary is negligible.

\section{Acknowledgements} F.P. acknowledges financial support through his Schr\"odinger Fellowship (Austrian Science Fund (FWF): J3675) at the University of Oxford and the NQIT project (EP/M013243/1). F.P. acknowledges travelling funding by the National University of Singapore for his visit in autumn $2015$ during which key ideas were communicated. X. R. was supported by Singapore Ministry of Education Academic Research Fund Tier 2 R-146-000-223-112.

We thank Weizhu Bao for assisting with the dynamical simulations in this paper and very helpful discussions. 

\section{Additional Information}

There are no competing financial interests.

\hspace{10mm}

F.P. wrote the main manuscript text and developed the analysis and the concept, T.A and R.X. developed numerics and R.X. prepared figures. All authors reviewed the manuscript.

 \end{document}